\documentclass[12pt]{article} 
\pdfoutput=1
\usepackage{amssymb,graphicx}
\usepackage{epstopdf}
\usepackage{amsmath,amsfonts}
\usepackage{epsfig} 
\usepackage[dvipsnames]{xcolor}
\usepackage{graphicx,graphics}
\usepackage[colorlinks=black,citecolor=blue,linkcolor=blue,breaklinks=true]{hyperref}

\begin{document} 

\sloppy

\title{\bf Gravitational waves from melting cosmic strings}
\author{William~T.~Emond, Sabir~Ramazanov, Rome~Samanta\\
\small{\em CEICO, FZU-Institute of Physics of the Czech Academy of Sciences,}\\
\small{\em Na Slovance 1999/2, 182 21 Prague 8, Czech Republic}
}

{\let\newpage\relax\maketitle}

{\let\newpage\relax\maketitle}

\begin{abstract}

Appearance of cosmic strings in the early Universe is a common manifestation of new physics typically linked to some high energy scale. In this paper, we discuss a different situation, where a model underlying cosmic string formation is approximately scale free. String tension is naturally related to the square of the temperature of the hot primordial plasma in such a setting, and hence decreases with (cosmic) time. With gravitational backreaction neglected, the dynamics 
of these melting strings in an expanding Universe is equivalent to the dynamics of constant tension strings in a Minkowski spacetime. 
We provide an estimate for the emission of gravitational waves from string loops. Contrary to the standard case, the resulting spectrum is markedly non-flat and has a characteristic falloff 
at frequencies below the peak one. The peak frequency is defined by the underlying model and lies in the range accessible by the future detectors for very weak couplings involved.

\end{abstract}

\section{Introduction}\label{sec:intro}

Searching for topological defects~\cite{Kibble:1976sj} in cosmological and astrophysical backgrounds is a promising way 
to probe physics beyond the Standard Model at very high energies inaccessible at Earth based facilities. In the present work, we focus on cosmic strings, which have a rich phenomenology, notably through their effect on the matter distribution 
of the Universe, gravitational lensing, and emission of gravitational waves (GWs)~\cite{Vilenkin, Hindmarsh:1994re}.  
The impact of cosmic strings can be described by one dimensionless parameter $G\mu$, where $G$ is Newton's constant and $\mu$ is a tension (mass density per unit mass length), commonly assumed to be time-independent. 
Consistency with the Planck data implies the limit $G\mu \lesssim 1.5 \cdot 10^{-7}$ at $95\%$ CL~\cite{Planck:2013mgr}. A considerably stronger constraint comes from pulsar timing arrays (PTAs), which reads $G \mu \lesssim 1.5 \cdot 10^{-11}$ at $95\%$ CL~\cite{Blanco-Pillado:2017rnf}. Note, however, that larger values of $G\mu$, which are in tension with this upper bound, are still of interest due to the recent results from NANOGrav~\cite{NANOGrav:2020bcs} hinting the first detection of stochastic GW background. This signal, if attributed to GWs, can be interpreted in terms of emission from cosmic strings~\cite{Ellis:2020ena, Samanta:2020cdk, Blasi:2020mfx}.

In this paper, we discuss a different type of cosmic strings, which have a tension decreasing with time\footnote{A weak logarithmic time-dependence of the tension is common for global cosmic strings. 
See~\cite{Chang:2019mza, Gorghetto:2021fsn} and references therein.}. The appearance of cosmic strings with such a seemingly exotic property can be well motivated in a physical setup, which involves no constant mass scale apart from the Planck scale due to a minimal coupling to gravity. We assume that the approximate scale-invariance is not spoiled by loop corrections. Consequently, the cosmic string tension should be related to the Hubble rate $H$, or the temperature of the Universe $T$. We choose the latter option, i.e., $\mu \propto T^2$, 
so that the time-dependence of the tension is fixed to be\footnote{The former option with $\mu \sim H^2$ has been considered in Ref.~\cite{Bettoni:2018pbl}.} 
\begin{equation}
\label{decreasing}
\mu \propto1/a^2 \; ,
\end{equation}
where $a$ is the scale factor of the Universe. A concrete example of a renormalizable model leading to such a behaviour is described in \hyperref[sec:Section2]{Section~2}. 
 
Evolution of cosmic strings is considerably simpler in scenario~\eqref{decreasing} compared to the case with constant $\mu$. Due to the scale-invariance of the model we consider, the dynamics of melting strings in an expanding Universe is equivalent to those of strings with a constant tension in a flat spacetime. We further elaborate on this in~\hyperref[sec:Section3]{Section~3} and in~\hyperref[sec:AppA]{Appendix~A} for the case of Nambu-Goto strings. This equivalence plays a crucial role for defining the number density of loops in the scaling 
regime: we simply use the results known from the studies of string evolution in a flat spacetime.
 
An interesting feature of scenario~\eqref{decreasing} is that one can have a large tension $\mu$ without conflicting with Cosmic Microwave Background (CMB) measurements. Indeed, even if $\mu$ starts from Planckian order values in the early Universe, 
it redshifts to a negligible value by recombination. Nevertheless, due to gravitational radiation emitted by the loops~\cite{Vachaspati:1984gt}, melting cosmic strings do not disappear without leaving a trace. 
In the present work, we estimate the GW spectrum in scenario~\eqref{decreasing} and show that it has a markedly non-flat shape. This contrasts the case of constant tension strings, cf. Fig.~\ref{gw}. In particular, the spectrum behaves 
as $\Omega_{gw} \propto f^4$ in the low frequency regime, which is directly related to the behaviour~\eqref{decreasing} and hence serves as a defining property of our scenario. 

In the approximation of Nambu-Goto strings and assuming that the particle emission by string loops is negligible, we estimate the present day fractional 
energy density of GWs, which can be as large as $\Omega_{gw} \simeq 10^{-8}-10^{-9}$ for $G\mu \simeq 10^{-4}$ at formation. Such energetic GWs are in the range accessible by essentially all planned detectors, provided 
that the peak frequency $f_{\text{peak}} \lesssim 100~\mbox{Hz}$.  The peak frequency is determined by the underlying model of melting strings. In the specific example discussed in \hyperref[sec:Section2]{Section~2}, 
phenomenologically interesting values of $f_{\text{peak}}$ imply extremely weak couplings of the fields constituting cosmic strings, to the thermal bath. In this regard, GW emission 
from melting strings may serve as a window to a (very) dark sector of the Universe. We comment on the Dark Matter implications of our scenario in \hyperref[sec:Section6]{Section~6}.

\section{From scale-invariance to melting cosmic strings} \label{sec:Section2}

In this section, we shall demonstrate via a specific model, how cosmic strings with the tension~\eqref{decreasing} arise. With this in mind, let us consider the following scale free renormalizable Lagrangian:
\begin{equation}
\label{modelbasic}
{\cal L}=-\frac{1}{4}F^{\mu \nu}F_{\mu \nu}+ \frac{1}{2} |D_{\mu} \chi|^2-\frac{1}{4}\lambda |\chi |^4 +\frac{1}{2} g^2 \cdot |\chi|^2 \cdot |\phi|^2 \; .
\end{equation}
Here $\chi$ is a field transforming under the $U(1)$ gauge group; $F_{\mu \nu} \equiv \partial_{\mu} A_{\nu} -\partial_{\nu} A_{\mu}$ is the gauge field $A_{\mu}$ strength tensor. The covariant derivative is given by $D_{\mu} \chi=\partial_{\mu} \chi-i e A_{\mu} \chi$, where $e$ is the gauge coupling constant. The field $\phi$ is a scalar multiplet comprising $N$ degrees of freedom. 

Assuming that $\phi$ is in thermal equilibrium with the surrounding plasma described by the temperature $T$, we fix its variance to be
\begin{equation}
\langle \phi^{\dagger} \phi \rangle_T \approx \frac{N T^2}{12} \; .
\end{equation}
Crucially we assume that $g^2$ is positive: 
\begin{equation}
\label{choice}
g^2>0 \; .
\end{equation}
As a result, the effective potential of the field $\chi$ has non-trivial minima $\chi_{min} \neq 0$ located at 
\begin{equation}
\label{minimum}
v \equiv |\chi_{\text{min}}| \approx \frac{g N^{1/2}T}{\sqrt{12 \lambda}} \; .
\end{equation}
Note that the model~\eqref{modelbasic} with the choice of the sign as in Eq.~\eqref{choice} was discussed in Ref.~\cite{Ramazanov:2021eya}, 
and we could readily use some of the results derived there. In particular, following Ref.~\cite{Ramazanov:2021eya} one can consider the field $\chi$ for the role 
of Dark Matter. We briefly discuss this option in \hyperref[sec:Section6]{Section~6}. However, the scenario of Ref.~\cite{Ramazanov:2021eya} deals with the $Z_2$-symmetry group, thus leading to melting domain walls instead of cosmic strings and hence to quite distinct predictions regarding GWs.

We assume that initially the field $\chi$ is at zero, $\chi \simeq 0$, and remains stuck there for some time due to Hubble friction. 
Rolling of the field $\chi$ to the minimum of the broken phase starts at the time $t_h$, when the Hubble rate becomes of the order of the thermal mass: 
\begin{equation}
\label{startroll}
|M_{\text{thermal}, h}| \approx \frac{gN^{1/2}T_h }{\sqrt{12}} \simeq H_h \; . 
\end{equation}
As the field $\chi$ reaches the minimum, a cosmic string network starts to form. Note that the rolling phase has a finite duration, and as a result, formation of cosmic strings is postponed until the time $t_l > t_h$. To capture this delay, we introduce the parameter
\begin{equation}
\label{epsilon}
\epsilon \equiv \frac{a_h}{a_l} \approx \frac{T_l}{T_h}  \; ,
\end{equation}
which is independent of $g$ and $\lambda$, and depends on the quantum fluctuation of the field $\chi$ above the background $\chi=0$. 
This quantum fluctuation, defined by the past evolution of the field $\chi$ at inflation and reheating, is crucial in triggering the roll towards the minimum. 
It is worth remarking here, that the parameter $\epsilon$ naturally takes values in the range $\epsilon \simeq 0.1-1$.

We assume that the transition to the spontaneously broken phase occurs at the radiation-dominated stage, so that
\begin{equation}
\label{Hubble}
H (T) =\sqrt{\frac{\pi^2 g_* (T)}{90}} \cdot \frac{T^2}{M_{\text{Pl}}} \; ,
\end{equation} 
where $g_* (T)$ is the number of relativistic degrees of freedom; $M_{\text{Pl}} \approx 2.44 \cdot 10^{18}~\mbox{GeV}$ is the reduced Planck mass.
Combining Eqs.~\eqref{startroll},~\eqref{epsilon}, and~\eqref{Hubble}, one obtains the temperature at the onset of cosmic string formation:
\begin{equation}
\label{formation}
T_{l} \simeq 9 \cdot 10^{-2} \cdot \epsilon \cdot g \cdot M_{Pl} \cdot N^{1/2} \cdot \left(\frac{100}{g_* (T_l)} \right)^{1/2} \; .
\end{equation}
Substituting this into Eq.~\eqref{minimum}, we get the expectation value at the time $t_l$:
\begin{equation}
\label{expectation}
v_{l} \approx \frac{2.6 \cdot 10^{-2} \cdot \epsilon  \cdot M_{Pl} \cdot N}{\sqrt{\beta}} \left(\frac{100}{g_* (T_l)} \right)^{1/2}\; ,
\end{equation}
where $\beta$ is defined as 
\begin{equation}
\label{beta}
\beta \equiv \frac{\lambda}{g^4} \; .
\end{equation}
The minimal possible value of $\beta$ follows from stability in the $(\chi, \phi)$ field space, i.e., $\lambda \lambda_{\phi} \geq g^4$, where $\lambda_{\phi}$ is the self-interaction coupling constant of the field $\phi$. Consequently, $\beta$ is bounded as
~\cite{Ramazanov:2021eya}  
\begin{equation}
\beta \geq \frac{1}{\lambda_{\phi}} \gtrsim 1\; .
\end{equation}
The latter inequality guarantees that we are in a weakly coupled regime, so that $\lambda_{\phi} \lesssim 1$. 
Furthermore, the condition $\beta \gtrsim 1$ ensures smallness of loop quantum corrections, $\delta \lambda \simeq N g^4/ (4\pi^2)$. 
In what follows, we will be primarily interested in very small self-interaction coupling constants $\lambda \simeq g^4$, corresponding to $\beta \simeq 1$. 
Such a choice is natural, if the model enjoys an approximate shift-invariance, becoming an exact one in the limit $g \rightarrow 0$.

The cosmic string tension is primarily defined by the expectation value $v$: 
\begin{equation}
\mu =\pi v^2 h \left(\frac{\lambda}{2e^2} \right)\; ,
\end{equation}
where $h$ is a slowly varying function of its argument $\lambda/(2e^2)$ (see below). According to Eq.~\eqref{minimum}, the tension relies on the square of the Universe temperature, $\mu \propto T^2$, and hence decreases with time as $\mu \propto 1/a^2$. Using Eq.~\eqref{expectation}, we obtain for the relevant quantity $G\mu$ at cosmic string formation:
\begin{equation}
\label{atformation}
G \mu_{l} \approx \frac{0.8 \cdot \epsilon^2 \cdot 10^{-4} \cdot N^2 \cdot h \left(\frac{\lambda}{2e^2} \right)}{\beta} \cdot \left(\frac{100}{g_* (T_{l})} \right) \; .
\end{equation}
For $g_* (T_l) \simeq 100-1000$, $N \simeq 1-10$, $\epsilon \simeq 0.1-1$, and $h \simeq 0.1-1$, the quantity $G\mu_l$ varies in the range
\begin{equation}
\label{range}
G\mu_{l} \simeq \frac{(10^{-8}-10^{-2})}{\beta} \; .
\end{equation}
Following the discussion above, we choose $\beta \simeq 1$ meaning that $G\mu_l \gtrsim 10^{-8}$. Intriguingly, in our setup, even cosmic strings with $G\mu_l \gg 10^{-7}$ are harmless for the CMB, because the decreasing tension becomes negligibly small at recombination. On the other hand, early time emission of GWs, when $\mu$ is large, can be detectable by the future GW interferometers or PTAs. 

Our choice $\lambda \simeq g^4$ imposes an important limitation on the gauge coupling constant $e$. Indeed, the Coleman-Weinberg one loop correction to the effective potential is given by
\begin{equation}
V_{CW} \simeq \frac{3e^4}{4 \pi^2} |\chi|^4 \ln \frac{|\chi|^2}{\sigma^2} \; ,
\end{equation}
where $\sigma$ is the renormalization scale. Note that the Coleman-Weinberg correction itself may lead to the spontaneous breaking of the $U(1)$ symmetry. In that case, one expects formation of cosmic strings with 
a constant tension. In the present work, we are interested in a different scenario, where quantum effects do not spoil scale-invariance of the model. That is, we require that the Coleman-Weinberg correction gives a negligible contribution to the effective potential of the field $\chi$, in particular, to the self-interaction term $\sim \lambda |\chi|^4$. Consistency with $\lambda \simeq g^4$ then implies that $e \lesssim g$. For $\lambda \simeq 2e^2$, one has $h \simeq 1$, while for $\lambda \ll 2e^2$ the following asymptotic behaviour holds~\cite{Hill:1987qx}:
\begin{equation}
\label{hlarge}
h \left(\frac{\lambda}{2e^2} \right) \simeq  \left(\ln \frac{2e^2}{\lambda} \right)^{-1}  \; .
\end{equation}
This explains the range of values $h \simeq 0.1-1$ assumed in Eq.~\eqref{range}\footnote{For $e \ll \sqrt{\lambda}$, one has $h \simeq \ln \frac{\lambda}{2e^2} \gg 1$. However, in this case gauge bosons 
are very light. As a result, most of the energy of the long string network goes into gauge field excitations rather than into loops, by analogy with the case of global strings. Reduced loop formation implies 
less gravitational radiation emitted compared to the case of larger gauge couplings $e \gtrsim \sqrt{\lambda}$.}.

Two important remarks are in order here. Generically, the motion of cosmic strings travelling in the surrounding plasma is slowed down by the thermal friction~\cite{Vilenkin}. Contrary to the case of constant tension cosmic strings, in our setup the thermal friction rate estimated as $\Gamma_{thermal} \sim T^3/\mu$ grows with time relative to the Hubble rate $H \sim T^2/M_{Pl}$. Therefore, one expects the motion of melting cosmic strings to be completely damped eventually. 
Naively, this should compromise any phenomenological applications of the model, in particular GW emission. 
However, the thermal friction can be neglected in the following analysis, if it is very small initially and comes into play only after a large number of Hubble times. Indeed, most energetic GWs 
are emitted close to the times of cosmic string formation in our case, and thus the late time evolution of cosmic strings is irrelevant for GW phenomenology. The thermal friction is negligible provided that $\Gamma_{thermal} \ll H$, which translates into the constraint on the temperature:
\begin{equation}
T \ll 8\pi^2 \cdot (G\mu) \cdot M_{Pl} \cdot \left(\frac{g_* (T)}{100} \right)^{1/2} \; .
\end{equation}
 Then, substituting the values of the temperature~\eqref{formation} at the cosmic string formation, we obtain the limit on the portal constant: 
 \begin{equation}
 g \ll \frac{800\pi^2}{N^{1/2}} \cdot \left(\frac{g_* (T_l)}{100} \right)  \cdot \left(\frac{0.1}{\epsilon} \right) \cdot (G\mu_l)  \; .
 \end{equation}
We will see in Section~5 that this is a very weak constraint in a sense that it does not impose any limitation on the portion of the parameter space, where one expects observable GWs. Therefore, we ignore the thermal friction in what follows.

Our second remark concerns the width $\delta_{w}$ of melting cosmic strings. The width $\delta_{w}$ always constitutes a small fraction of the horizon for $\epsilon \ll 1$ (cf. Ref.~\cite{Bettoni:2018pbl}), i.e.,
\begin{equation}
\label{width}
\delta_{w} H \simeq \epsilon \cdot \frac{a_l}{a(t)} \; .
\end{equation}  
The time dependence here follows from $\delta_{w} \simeq 1/{\sqrt{\lambda} v} \propto a(t)$ and $H \propto 1/a^2(t)$. The normalization factor in Eq.~\eqref{width} is obtained by combining Eqs.~\eqref{Hubble},~\eqref{formation}, and~\eqref{expectation}. 
We see that the cosmic string width, while shrinking with time relative to the Hubble radius, initially constitutes a considerable fraction of the distance between infinite strings as well as the size of the largest loops. Since the early times, close to the 
moment of the network formation $t_i$, are most relevant from the viewpoint of GW phenomenology discussed in \hyperref[sec:Section4]{Section~4}, melting cosmic strings should be treated as Abelian-Higgs rather than Nambu-Goto ones. Nevertheless, in the following analysis 
we often rely on expressions and results of numerical simulations obtained assuming Nambu-Goto strings. We assume that this is plausible for the purpose of crude estimation in most cases, and we comment on the situations where we expect 
departures from the approximation of Nambu-Goto strings to be unacceptably large.

\section{Number density of cosmic string loops} \label{sec:Section3}
The evolution of cosmic strings in the model~\eqref{modelbasic} is greatly simplified as a result of its scale-invariance. Indeed, by carrying out the following field redefinitions of the variables $\chi$ and $A_\mu$:
\begin{equation}
\label{redefined}
\tilde{\chi}=\chi a \qquad \tilde{A}_{\mu}=A_{\mu}\; ,
\end{equation}
we find that cosmic strings within this model are described by the action 
\begin{equation}
S=\int d^{3} {\bf x} d \tau \left[-\frac{1}{4} \eta^{\lambda \mu} \eta^{\rho \nu} F_{\lambda \rho} F_{\mu \nu}+\frac{1}{2} \eta^{\mu \nu} D_{\mu} \tilde{\chi} D_{\nu} \tilde{\chi}^{*}-\frac{\lambda}{4} \left(|\tilde{\chi}|^2 -\tilde{v}^2 \right)^2 \right] \; ,
\end{equation}
where $\tau$ is the conformal time, $\eta^{\mu \nu}$ is the Minkowski metric, and $\tilde{v}$ is the expectation value of the field $\tilde{\chi}$, i.e., $\tilde{v}=v \cdot a$. Hereafter, rescaled quantities are denoted by a tilde. According to Eq.~\eqref{minimum}, $v \propto 1/a$, hence $\tilde{v}$ is constant. We conclude that evolution of melting cosmic strings in a radiation-dominated Universe is equivalent to evolution of constant tension cosmic strings in Minkowski spacetime. In \hyperref[sec:AppA]{Appendix~A}, we reiterate this statement starting from the Nambu--Goto action. There we also discuss the generalization to cosmic strings whose tension has arbitrary time dependence.

The cosmic string network enters the scaling regime soon after its formation. In the case of flat spacetime, this has been demonstrated with numerical simulations in Refs.~\cite{Vanchurin:2005yb} (for long strings) and~\cite{Vanchurin:2005pa} (for loops). In the scaling regime, the loop production 
function--the number density of loops per unit conformal time per unit loop conformal length $\tilde{l} \equiv \frac{l(t)}{a(t)}$--is given by 
\begin{equation}
f (\tau, \tilde{l}) =\frac{1}{\tau^5} f (x) \; ,
\end{equation}
where 
\begin{equation}
x \equiv \frac{\tilde{l}}{\tau} \; .
\end{equation}
The form of $f(x)$ is independent of time. 
The number density of loops per unit length produced in the scaling regime is related to the loop production function by  
\begin{equation}
n (\tau, \tilde{l}) \equiv \frac{dN}{d{\bf x} d\tilde{l}}=\int^{\tau}_{\tau_{s}} \frac{d\tau'}{\tau^{'5}} f (x') \; ,
\end{equation}
where $\tau_s$ denotes the time, when the cosmic string network settles into the scaling regime. Neglecting gravitational backreaction, the loop length $\tilde{l} =x \tau$ remains constant with time. Therefore, one has 
\begin{equation}
x' \tau' =x \tau.
\end{equation}
Following Ref.~\cite{Blanco-Pillado:2013qja}, we make the change of the integration variable: 
\begin{equation}
n (\tau, \tilde{l}) =\int^{\tilde{l}/\tau}_{\tilde{l}/\tau_{s}} \frac{dx'}{\tau^{'5}}  f (x') \frac{\partial \tau'}{\partial x'}     \; ,
\end{equation}
and obtain
\begin{equation}
\label{flatscaling}
n (\tau, \tilde{l})=\frac{1}{\tilde{l}^4} \int^{\tilde{l}/\tau_{s}}_{\tilde{l}/\tau} dx' x^{'3} f(x')  \; .
\end{equation}
The loop number density per comoving volume per physical length $n(t, l)$ is related to $n (\tau, \tilde{l})$ by 
\begin{equation}
\label{rel}
n (t, l) \equiv \frac{dN}{d{\bf x} dl}= \frac{1}{a(\tau)} n (\tau, \tilde{l}) \; .
\end{equation}
In the limit $\tau_{s} \rightarrow 0$, the integral in Eq.~\eqref{flatscaling} takes the form of Eq.~(15) in Ref.~\cite{Blanco-Pillado:2013qja}\footnote{One should also set $\nu=0$ in Eq.~(15) of Ref.~\cite{Blanco-Pillado:2013qja}, which corresponds to the 
flat spacetime limit.}. However, as we will see in \hyperref[sec:Section4]{Section~4}, the most relevant contribution to GW emission comes from very early times. Therefore, it is crucial that we do not take the limit $\tau_s \rightarrow 0$.

To proceed, we need to make a choice of the loop production function $f(x)$. When evaluating GWs, we cannot reliably account for contributions arising from small loops, with $x \ll 0.1$, for two main reasons. First, in the model described in \hyperref[sec:Section2]{Section~2}, strings are relatively thick at the time of production, 
and thus we expect small loop formation to be suppressed. Second, small loops, if abundantly produced, initially exhibit a strong departure from the scaling regime, which persists for a longer time compared to large loops~\cite{Vanchurin:2005pa, Ringeval:2005kr, Martins:2005es, Olum:2006ix}. Therefore, 
we mostly discard small loops with $x \ll 0.1$ in the following analysis, possibly at the price of underestimating gravitational radiation. In this situation, it is natural to stick to the velocity-dependent one-scale model (VOS)~\cite{Kibble:1984hp, Bennett:1985qt}, which assumes that the size of loops created at any time $\tau$ 
is a fixed fraction of $\tilde{L} \propto \tau$ corresponding to the distance between long strings. 
This size is chosen to match the maximum size of loops obtained in numerical simulations~\cite{Vanchurin:2005pa}: $\tilde{l} =\alpha \tau$, where 
\begin{equation}
\label{maximum}
\alpha \simeq 0.1 \; .
\end{equation} 
That is, the function $f(x)$ is given by 
\begin{equation}
\label{onescale}
f(x)=C \delta \left(x-\alpha \right) \; .
\end{equation} 
There are two ways of getting the constant $C$ -- one is analytical and is summarized in \hyperref[sec:AppB]{Appendix~B}. It gives $C \simeq 500$.
The other involves matching to the function $f(x)$ derived from numerical simulations of cosmic strings in a flat spacetime~\cite{Vanchurin:2005pa}: 
\begin{equation}
\label{loopdensity}
f (x) \approx \frac{A \Theta \left(\alpha-x \right)}{x^{\gamma}}\; ,
\end{equation}
where $A \approx 82$ and $\gamma \approx 1.63$. The constant $C$ is fixed by the requirement that Eqs.~\eqref{onescale} and~\eqref{loopdensity} give the same number density of loops, when substituted into Eq.~\eqref{flatscaling}:
\begin{equation}
\label{numerics}
C \approx 150 \; .
\end{equation} 
This is only a factor three below the analytically derived value. We shall assume the value~\eqref{numerics} in the following analysis.

Due to emission of GWs, loops shrink in size as time proceeds, and this effect should be taken into account when defining the number density of loops. However, as we shall now demonstrate, gravitational backreaction practically does not affect the evolution of large loops in our case. The mass of the loop $m \equiv \mu l$ is changing according to 
\begin{equation}
\frac{dm}{dt}= -H m -\Gamma G\mu^2 (t)\; ,
\end{equation}
where $\Gamma \approx 50$~\cite{Vachaspati:1984gt, Blanco-Pillado:2017oxo}. The first term on the R.H.S. follows from an approximate scale-invariance of the model: all quantities with mass dimension redshift with the scale factor as $m \sim 1/a$ (modulo gravitational backreaction). Consequently, evolution 
of the loop conformal length is given by 
\begin{equation}
\frac{d\tilde{l}}{d\tau}=-\frac{\Gamma G \tilde{\mu}}{a^2 (\tau)} \; .
\end{equation}
Integrating this out, we get
\begin{equation}
\tilde{l} (\tau')=\tilde{l} (\tau) - \frac{\Gamma G \tilde{\mu} \cdot \tau}{a^2 (\tau)} \cdot \left(1-\frac{\tau}{\tau'} \right) \; .
\end{equation}
We see that for any given $\tau$ the second term on the R.H.S. reaches a constant value $\Gamma G \mu (\tau) \tau$ (recall that $\mu (\tau)=\tilde{\mu}/a^2 (\tau)$).
Thus, for $\tilde{l} (\tau) \gg  \Gamma G \mu (\tau) \tau$, one can neglect gravitational backreaction. As soon as we are interested in large loops with $\tilde{l}/\tau \simeq 0.1$, this inequality is fulfilled, at least if the initial tension is not too large, $G \mu_l \lesssim 10^{-3}$. On the contrary, in the case of constant tension cosmic strings, gravitational backreaction is being accumulated with time, so that loops evolve according to $l (t') =l (t)+\Gamma G \mu (t'-t)$, hence any loop evaporates at some point. 

Generically cosmic strings also emit particles (gauge and scalar bosons) on top of gravitational radiation. The lower bound on the length of loops, which predominantly radiate GWs is estimated as (see Eq.~(11) in Ref.~\cite{Matsunami:2019fss}):
 \begin{equation}
 \label{particletrue}
 l \gtrsim \frac{\epsilon_p}{G\mu^2} \; ,
 \end{equation}
 where $\epsilon_p$ is the energy of particle radiation produced per an emission episode. For $\lambda \sim e \sim 1$ assumed in Refs.~\cite{Matsunami:2019fss, Auclair:2019jip}, one naturally estimates 
$\epsilon_p \sim \mu^{1/2}$ and $\delta_{w} \sim \mu^{-1/2}$, and the inequality~\eqref{particletrue} reduces to $l \gtrsim \frac{\delta_{w}}{G\mu}$ confirmed by running lattice simulations; thus, sufficiently thick strings predominantly decay into particles. This conclusion is not applicable to our case, because we deal with a very different 
parameter space compared to Refs.~\cite{Matsunami:2019fss, Auclair:2019jip}, i.e., $\lambda~\sim g^4 \ll 1$ and $e \lesssim g \ll 1$. While the string width is estimated as $\delta_{w} \sim 1/\sqrt{\lambda \mu}$ (for $e \gtrsim \sqrt{\lambda}$), we do not know, how to estimate $\epsilon_p$ for generic $\lambda$ and $e$. According to Eq.~\eqref{particletrue}, particle emission is negligible at least for large loops with the length $l \sim 0.1~H^{-1}$, provided that
\begin{equation}
\label{assumption}
\epsilon_p \ll (G\mu) \cdot \sqrt{\frac{\mu}{\lambda}} \; .
\end{equation}
In the remainder of this paper, we assume that Eq.~\eqref{assumption} is fulfilled, 
so that particle emission of cosmic string loops is negligible compared to GW emission.

\section{Gravitational waves from cosmic strings} \label{sec:Section4}

\subsection{Generalities}

Cosmic string loops with a given length $l(t)$ emit GWs at frequencies $F^{(j)}=2j/l(t)$, where $j=1,2,3,...$ is the multipole number~\cite{Vachaspati:1984gt}.  
These frequencies redshift with cosmic expansion, so that their current values are given by $f^{(j)}=(a(t)/a_0) \cdot F^{(j)}$, or
\begin{equation}
f^{(j)} =\frac{a(t)}{a_0} \cdot \frac{2j}{l (t)}=\frac{2j}{a_0 \tilde{l}} \; .
\end{equation}
Our goal is to obtain the fractional energy density of GWs emitted by loops
\begin{equation}
\Omega_{gw} (f) \equiv \frac{fd\rho_{gw}}{\rho_c d f} \; ,
\end{equation}
where $\rho_{gw}$ is the present day energy density of GWs and $\rho_c$ is the critical energy density. 
We split $\Omega_{gw} (f)$ into a sum over harmonics: 
\begin{equation}
\Omega_{gw} (f)=\sum^{\infty}_{j=1} \Omega^{(j)}_{gw}=\sum^{\infty}_{j=1}\frac{fd\rho^{(j)}_{gw}}{\rho_c d f} \; .
\end{equation}
Following Ref.~\cite{Blanco-Pillado:2013qja}, we relate $\frac{d\rho^{(f)}_{gw}}{df}$ to the power of GW emission per unit physical volume per unit frequency ${\cal P}^{(j)}_{gw} (t, F)$:
\begin{equation}
\frac{d\rho^{(j)}_{gw}}{d f} =\int^{t_0}_{t_{l}} dt'  \left(\frac{a (t')}{a_0} \right)^4 \cdot {\cal P}^{(j)}_{gw} (t', F') \cdot \frac{\partial F'}{\partial f}=\int^{t_0}_{t_l} dt' \left(\frac{a(t')}{a_0} \right)^3 \cdot {\cal P}^{(j)}_{gw}  \left( t', \frac{a_0}{a(t')} f \right) \; .
\end{equation}
Here the factors $(a(t')/a_0)^4$ and $\partial F'/\partial f=a_0/a(t')$ take into account the redshift of the GW energy density and frequency, respectively. In turn, ${\cal P}^{(j)}_{gw}$ is defined by the comoving number density of loops per unit length $n(t, l)$ through
\begin{equation}
{\cal P}^{(j)}_{gw} (t, F)=\int dl \frac{n (t, l)}{a^3 (t)} P^{(j)}(l ,F) \; ,
\end{equation}
where $P^{(j)}(l ,F)$ is the power emitted by a single loop. Assuming that loops develop and maintain only one cusp, one has in the approximation of Nambu-Goto strings~\cite{Vachaspati:1984gt, Blanco-Pillado:2013qja, Blanco-Pillado:2017oxo}\footnote{The sub-dominant contributions to the GW power come from kinks and kink-kink collisions; they decay as $j^{-5/3}$ and $j^{-2}$, respectively.}
\begin{equation}
\label{singleloopgw}
P^{(j)}_{gw} (l, F)=\frac{\Gamma G \mu^2 (t)}{\zeta \left(\frac{4}{3}, \infty \right)} \frac{1}{j^{4/3}} \delta \left(F- \frac{2j}{l (t)}  \right) \; ,
\end{equation}
where 
\begin{equation}
\zeta \left(\frac{4}{3}, \infty \right) =\sum^{\infty}_{j=1} \frac{1}{j^{4/3}} \approx 3.60 \; .
\end{equation}
Strictly speaking the formula~\eqref{singleloopgw} works only for $j \gg 1$, 
however, the deviation from Eq.~\eqref{singleloopgw} at $j \sim 1$ is within a factor of two~\cite{Blanco-Pillado:2017oxo}, and we ignore it in the following. 
Note that Eq.~\eqref{singleloopgw} is commonly written for a constant tension, nevertheless, the extrapolation to a time-independent tension is legitimate, as soon as it changes 
negligibly during an oscillation period. This is indeed the case here, as $F^{(j)} \gg H$. 

As melting cosmic strings have a sizeable width (see the last paragraph of \hyperref[sec:Section2]{Section~2}), it is important to discuss to which extent using Eq.~\eqref{singleloopgw} is trustworthy. 
In the case of infinitely thin cosmic strings the range for the multipole number $j$ extends to infinity. In our case, the frequency of GWs $F^{(j)}$ exceeds $\delta^{-1}_{w}$ for $j \gtrsim l/\delta_{w}$, and departures from the approximation of Nambu-Goto strings become unacceptably large. 
We expect the inverse string width to set the cutoff scale on the frequencies of emitted GWs meaning a strong reduction of the power compared to Eq.~\eqref{singleloopgw}. Therefore, we limit to the values $j$ close to unity in what follows.

Combining everything altogether, we get 
\begin{equation}
\label{gen}
\Omega^{(j)}_{gw} (f)=\frac{1}{\rho_c}  \cdot \frac{2\Gamma G}{\zeta \left(\frac{4}{3}, \infty \right) j^{1/3} \cdot f} \cdot \int^{t_0}_{t_{l}} dt' \frac{a^2 (t') \mu^2 (t') n\left(t', \frac{2j a(t')}{a_0f} \right) }{a^5_0} \; .
\end{equation}
To proceed, we make an assumption that the scaling regime is reached almost immediately upon string formation, i.e., 
\begin{equation}
\label{instantscaling}
t_{l} \simeq t_{s} \; .
\end{equation}
In reality, it takes some time for the network to settle down into the scaling behaviour. We will see, however, that some part of GW spectrum, namely the low-frequency part, is unaffected 
by the early time departure from the scaling regime. For the rest of the spectrum we will be satisfied with a crude estimate. With that said, we use Eqs.~\eqref{flatscaling} and~\eqref{rel} to define the number density of loops entering Eq.~\eqref{gen}:
\begin{equation}
n\left(t, \frac{2j a(t)}{a_0f} \right) = \frac{1}{a(\tau)} n\left(\tau, \frac{2j}{a_0f} \right) \simeq \frac{a^4_0 f^4}{16 a(\tau) j^4} \int^{\frac{2j}{a_0f \tau_{l}}}_{\frac{2j}{a_0f \tau}} dx x^3 f(x) \; .
\end{equation}
Finally, switching to conformal time and using $\rho_c =3H^2_0 M^2_{Pl}$, where $H_0$ is the Hubble constant, we rewrite Eq.~\eqref{gen} as
\begin{equation}
\label{generic}
\Omega^{(j)}_{gw} (f) \simeq \frac{\pi \Gamma G^2 f^3}{3H^2_0 \zeta \left(\frac{4}{3}, \infty \right) j^{13/3}} \cdot \int^{\tau_0}_{\tau_{l}} d\tau' \frac{a^2 (\tau')}{a_0} \mu^2 (\tau') \int^{\frac{2j}{a_0 f \tau_{l}}}_{\frac{2j}{a_0 f \tau'}} dx' x^{'3} f (x') \; .
\end{equation}
Note that given the time-dependence of the tension $\mu$, the integral over $\tau'$ is saturated at very early times close to $\tau_l$. This is in contrast to the case of cosmic strings with a constant tension. 
In particular, this results into a markedly non-flat spectrum, as we discuss in details in the remainder of this section.

\subsection{Spectrum of gravitational waves} 
We follow the VOS approach to model the loop distribution, such that the function $f(x)$ is chosen to be of the form~\eqref{onescale}. In doing so, we explicitly neglect the contribution of small loops. As it has been mentioned in the previous sections, strings are rather thick initially, and therefore we expect production of small loops to be suppressed. Nevertheless, at the end of this section, we comment 
on the GW spectrum that would follow from the model~\eqref{loopdensity}, once an abundant production of small loops is assumed. We will see that the latter may strongly affect the spectrum in the intermediate frequency range, at the same time leaving intact the shape of the spectrum in the low- and high-frequency regimes.

{\it Low frequency range.} With the VOS model assumed, the characteristic frequency of GWs is given by 
\begin{equation}
\label{peakf}
f_{\text{peak}} \equiv  \frac{2}{a_0 \alpha \tau_{l}} \approx \frac{2 H_{l}}{\alpha} \cdot \frac{a_{l}}{a_0} \; .
\end{equation}
The fact that this is indeed the peak frequency will become clear shortly. 
We start with the low frequency regime, $f<f_{\text{peak}}$. In this case, using Eq.~\eqref{onescale}, we can express the inner integral in Eq.~\eqref{generic} as follows:
\begin{align}
\label{twooptions}
\int^{\frac{2j}{a_0 f \tau_{l}}}_{\frac{2j}{a_0 f \tau'}} dx' x^{'3} f (x')= 
\begin{cases}
0  ~\qquad \tau' < \tilde{\tau}^{(j)}\\ 
C \alpha^3 \quad \tau' \geq \tilde{\tau}^{(j)}\; ,
\end{cases}
\end{align}
where 
\begin{equation}
\label{latertime}
\tilde{\tau}^{(j)} \equiv \frac{2j }{a_0 \alpha f } = \tau_{l} \cdot j \cdot \frac{f_{\text{peak}}}{f} \; . 
\end{equation}
Substituting Eq.~\eqref{twooptions} into Eq.~\eqref{generic}, we obtain
\begin{equation}
\label{interexp}
\Omega^{(j)}_{gw} (f) \simeq\frac{\pi C\alpha^3 \Gamma G^2 f^3}{3H^2_0 \zeta \left(\frac{4}{3}, \infty \right) j^{13/3}} \cdot \frac{\mu^2 (\tilde{\tau}^{(j)}) a^2 (\tilde{\tau}^{(j)}) \tilde{\tau}^{(j)} }{a_0}   \; .
\end{equation}
We observe that $\mu (\tilde{\tau}^{(j)}) a^2 (\tilde{\tau}^{(j)}) =\mu_{l} \cdot a^2_{l}$ and $a (\tilde{\tau}^{(j)})=a_l \cdot (\tilde{\tau}^{(j)})/\tau_l$, and then use Eq.~\eqref{latertime} to express the time $\tilde{\tau}^{(j)}$ in terms of $\tau_l$. 
Next, we express the time $\tau_l$ through the Hubble rate $H_{l}=1/(a_{l} \tau_{l})$ and $\alpha$ through $f_{\text{peak}}$ by making use of Eq.~\eqref{peakf}. Finally, we get for $f<f_{\text{peak}}$:
\begin{equation}
\label{modesmall}
\Omega^{(j)}_{gw} (f) \simeq \frac{1}{j^{16/3}} \cdot \frac{8\pi C \Gamma  (G \mu_{l})^2}{3 \zeta \left(\frac{4}{3}, \infty \right)}  \cdot \left(\frac{H_{l}}{H_0} \right)^2 \cdot \left(\frac{a_{l}}{a_0} \right)^4 \cdot \left(\frac{f}{f_{\text{peak}}} \right)^4 \; .
\end{equation}
Clearly, higher multipoles with $j>1$ contribute negligibly to the low frequency range, so that the overall fractional energy density of GWs can be well approximated by counting only the fundamental 
harmonic $j=1$:
\begin{equation}
\label{low}
\Omega_{gw} (f<f_{\text{peak}}) \simeq \Omega_{gw, \text{peak}} \cdot \left(\frac{f}{f_{\text{peak}}} \right)^4 \; .
\end{equation}
Here $\Omega_{gw, \text{peak}}$ is the peak energy density of GWs given by
\begin{equation}
\label{peakgen}
\Omega_{gw, \text{peak}} \simeq \frac{8\pi C \Gamma (G \mu_{l})^2}{3 \zeta \left(\frac{4}{3}, \infty \right)}  \cdot \left(\frac{H_{l}}{H_0} \right)^2 \cdot \left(\frac{a_{l}}{a_0} \right)^4 \; .
\end{equation}
Using the values $\Gamma \approx 50$, $C \simeq 150$, $\zeta \left(\frac{4}{3}, \infty \right) \approx 3.60$, and 
\begin{equation}
\left(\frac{H_l}{H_0} \right)^2 \cdot \left(\frac{a_l}{a_0} \right)^4 \approx 2.6 \cdot 10^{-5} \cdot \left(\frac{100}{g_* (T_l)} \right)^{1/3} \; ,
\end{equation}
we obtain
\begin{equation}
\label{peak}
\Omega_{gw, \text{peak}} \simeq 4.5 \cdot 10^{-9} \cdot \left(\frac{G\mu_{l}}{10^{-4}} \right)^2 \cdot \left(\frac{100}{g_* (T_{l})} \right)^{1/3} \; .
\end{equation}
GWs emitted in the range of frequencies $f \ll f_{\text{peak}}$ provide a particularly clean probe of our scenario, because they come from times considerably later than $\tau_l$, when the departure from the scaling regime is moderate. One can also show, that the result~\eqref{low} is largely independent of $f(x)$. In particular, we would get the same had we chosen to use the loop production function an in Eq.~\eqref{loopdensity}.

 {\it High frequency range.} Now let us discuss the range of high frequencies with $f>f_{\text{peak}}$. In this case, the discussion above applies and in particular Eq.~\eqref{modesmall} holds, but only for multipole numbers $j > f/f_{\text{peak}}$. Indeed, for $j < f/f_{\text{peak}}$, one has 
 \begin{equation}
 \frac{2j}{a_0 f \tau_l} < \frac{2}{a_0 f_{\text{peak}} \tau_l} =\alpha 
 \end{equation}
 and hence, 
 \begin{equation}
 \int^{\frac{2j}{a_0 f \tau_l}}_{\frac{2j}{a_0f \tilde{\tau}^{(j)}}} dx' x^{'3} f(x') = C\int^{\frac{2j}{a_0 f \tau_l}}_{\frac{2j}{a_0f \tilde{\tau}^{(j)}}} dx' x^{'3} \delta \left(x'-\alpha \right) =0 \; .
 \end{equation}
 Thus, according to Eq.~\eqref{generic}, we must set $\Omega^{(j)}_{gw}=0$ for $j<f/f_{\text{peak}}$. Performing the summation in Eq.~\eqref{modesmall} from $j \approx f/f_{\text{peak}}$ and using Eq.~\eqref{peakgen} we obtain
 \begin{equation}
 \label{high}
 \Omega_{gw} (f > f_{\text{peak}}) \simeq c(f) \cdot \Omega_{gw, \text{peak}}  \left(\frac{f_{\text{peak}}}{f} \right)^{1/3} \; ,
 \end{equation}
 where $c(f)$ is defined as
 \begin{equation}
 \label{cf}
 c(f) =\left(\frac{f}{f_{\text{peak}}} \right)^{13/3} \cdot \sum^{\infty}_{j > f/f_{\text{peak}}} \frac{1}{j^{16/3}} \; .
 \end{equation}
The discontinuity between Eqs.~\eqref{low} and~\eqref{high} at $f=f_{\text{peak}}$ is a consequence of the one-scale approximation used. In reality, the function $f(x)$ is smooth around the peak $x \simeq \alpha$, and the discontinuity is avoided. 
One can check that the function $c(f)$ quickly relaxes to a constant value as one increases $f$: 
 \begin{equation}
 c(f) \rightarrow \frac{3}{13} \; .
 \end{equation}
In practice, we will set $c(f)=3/13$ for all $f>f_{\text{peak}}$. Note that the slow decay with the frequency in Eq.~\eqref{high} is independent of the choice of the loop production function $f(x)$, and merely reflects a generic dependence on the multipole number in Eq.~\eqref{singleloopgw}. As has been remarked on in \hyperref[sec:Section4]{Section~4}, using Eq.~\eqref{singleloopgw} is most probably not legitimate for $j \gg 1$ because of the issue with 
a finite string width, at least close to the cosmic string network formation time. Thus, extrapolating~\eqref{high} to
 large $f \gg f_{\text{peak}}$ is not legitimate, and we expect a strong falloff of the spectrum in a very high frequency range compared to Eq.~\eqref{high}.

{\it Possible contribution of small loops.} To estimate the possible effect of small loops 
neglected in the analysis above, one turns to the model~\eqref{loopdensity}. Again we assume that 
the string network settles immediately into the scaling regime upon its formation. The lower bound on the loop size is set by gravitational backreaction discussed at the end of \hyperref[sec:Section3]{Section~3}: $x \gtrsim \Gamma G \mu (\tau)$, where $x =\tilde{l}/\tau$. One can show 
that the loops saturating this lower bound give the main contribution to the peak energy density of GWs. As a result, the peak frequency of GWs is shifted compared to the VOS model:
\begin{equation}
f_{\text{peak}} \simeq f^{VOS}_{\text{peak}} \cdot \frac{\alpha}{\Gamma G \mu_l } \; , 
\end{equation}
where $f^{VOS}_{\text{peak}}$ is given by Eq.~\eqref{peakf}. In the intermediate frequency range $f^{VOS}_{\text{peak}}\lesssim f \lesssim f_{\text{peak}}$, there is a 
growth of GW energy density $\Omega_{gw} (f) \propto f^{\gamma-1}$. In particular, for the value $\gamma \approx 1.6$ 
fitting numerical simulations of Ref.~\cite{Vanchurin:2005pa}, one has $\Omega_{gw} (f) \propto f^{0.6}$. In other words, the inclusion of small loops leads to a larger value of $\Omega_{gw, \text{peak}}$ compared to Eq.~\eqref{peak}: 
\begin{equation}
\Omega_{gw, \text{peak}} \simeq \Omega^{VOS}_{gw, \text{peak}} \cdot \left(\frac{\alpha}{\Gamma G \mu_l} \right)^{0.6} \; .
\end{equation} 
Thus, Eq.~\eqref{peak} should be viewed as a conservative lower bound on the peak energy density of GWs (provided that the particle emission is negligible). On the other hand, the behaviour in the low and high frequency regimes, i.e.,  
$\Omega_{gw} \propto f^4$ for $f \lesssim f^{VOS}_{\text{peak}}$ and $\Omega_{gw} \propto f^{-1/3}$, for $f \gtrsim f_{\text{peak}}$, respectively, still holds.

\section{Prospects for observations}

\begin{figure}[tb!]
  \begin{center}
    \includegraphics[scale=.98]{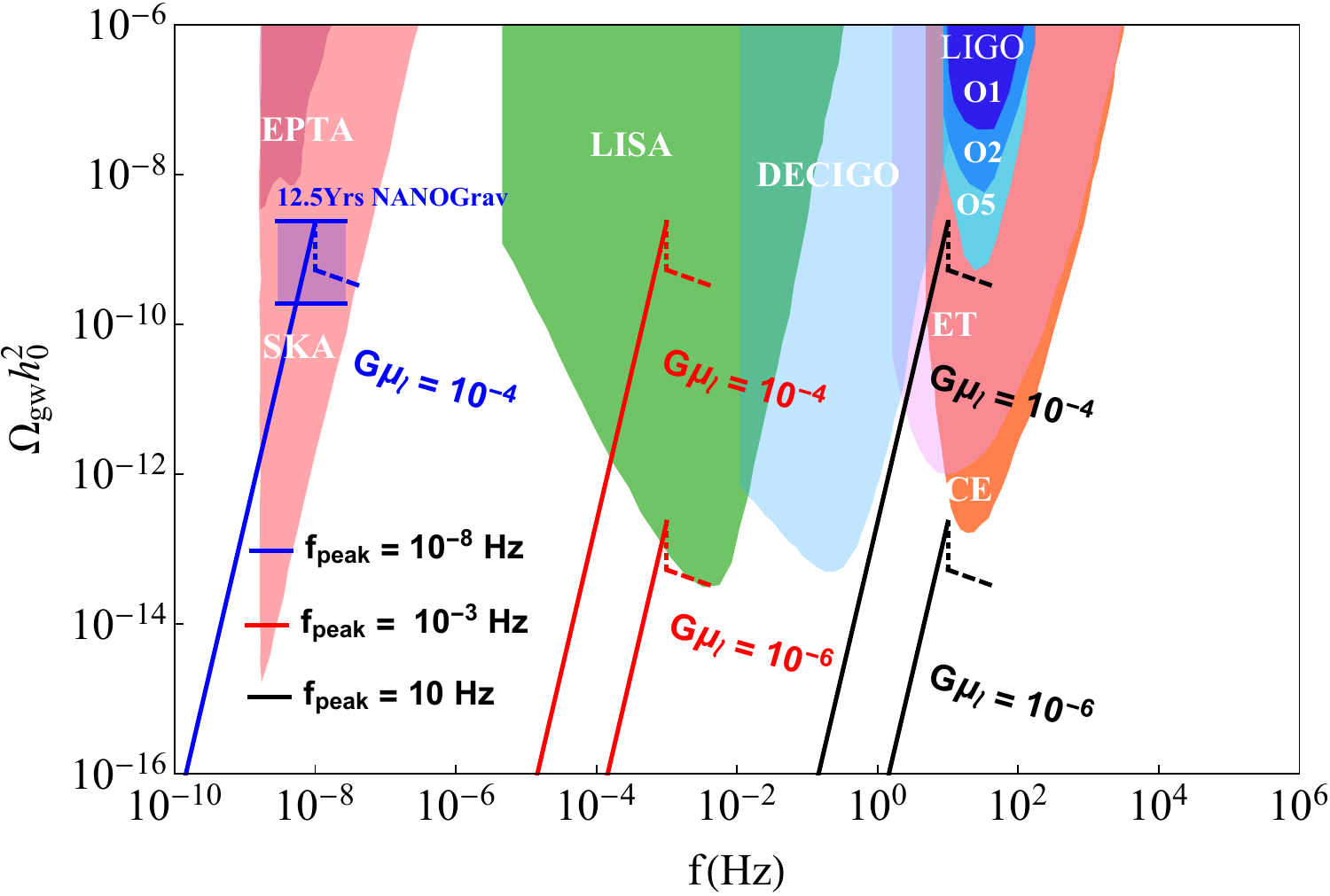}
  \caption{The spectrum of GWs emitted by the network of melting cosmic strings is shown for different values of $G\mu_l$ and $f_{\text{peak}}$. The choice of peak frequencies $f_{\text{peak}}$ corresponds to the choice of 
  portal constants $g \simeq 10^{-9},~10^{-13},~10^{-18}$ in the model~\eqref{modelbasic}. The shaded regions correspond to those accessible with current and future 
  GW detectors. Note that the spectrum of GWs was derived in the approximation of infinitely thin strings. Therefore, the high frequency part of the spectrum, where we expect the effects due to a finite string width to be 
  particularly large, is shown with dashed lines.}\label{gw}
  \end{center}
\end{figure}

In Fig.~\ref{gw}, we show the spectrum of GWs in terms of $\Omega_{gw} \cdot h^2_0$, where $h_0 \approx 0.7$ is the dimensionless Hubble constant, for different $G\mu_l$ and $f_{\text{peak}}$. We 
compare it with the sensitivity of current and future detectors: LIGO~\cite{LIGOScientific:2019vic, LIGOScientific:2014qfs}, Einstein Telescope (ET)~\cite{Sathyaprakash:2012jk}, Cosmic Explorer (CE)~\cite{LIGOScientific:2016wof}, DECIGO~\cite{Kawamura:2011zz}, LISA~\cite{LISA:2017pwj, Auclair:2019wcv}, and PTAs~\cite{NANOGrav:2020bcs, Kramer:2013kea, Janssen:2014dka}. For the peak 
frequency lying in the ranges $10^{-9}~\mbox{Hz} \lesssim f_{\text{peak}} \lesssim 10^{-7}~\mbox{Hz}$ and $10^{-5}~\mbox{Hz} \lesssim f_{\text{peak}} \lesssim 100~\mbox{Hz}$, one will be able to probe melting strings with an initial tension 
as large as $G\mu_l \gtrsim 10^{-6}$. Note that astrometrical measurements have a promising capability to fill in the existing frequency gap between the SKA and LISA sensitivity curves~\cite{Garcia-Bellido:2021zgu}. 

So far, our discussion of the GW spectrum has been model-independent: it only assumed that the tension decreases with time as $\mu \propto 1/a^2$. 
Now, let us relate the peak frequency $f_{\text{peak}}$ of GW emission given by Eq.~\eqref{peakf}  to the parameters of the model~\eqref{modelbasic}. One first connects $f_{\text{peak}}$ with the temperature $T_l$ at string formation: 
\begin{equation}
\label{peaktemperature}
f_{\text{peak}} \approx 20 \cdot \frac{T_l \cdot T_0}{M_{Pl}} \cdot \left(\frac{g_* (T_l)}{100} \right)^{1/6} \; ,
\end{equation}
where $T_0 \approx 2.73~\mbox{K}$ is the present day temperature of the Universe. Next we substitute Eq.~\eqref{formation} into Eq.~\eqref{peaktemperature} and obtain
\begin{equation}
\label{peakfreq}
f_{\text{peak}} \simeq 100~\mbox{Hz} \cdot \sqrt{N} \cdot \left(\frac{g}{10^{-8}} \right) \cdot \left(\frac{\epsilon}{0.1} \right) \cdot \left(\frac{100}{g_* (T_l)} \right)^{1/3} \; .
\end{equation}
Note that the peak frequency is mainly defined by the portal constant $g$ and is independent of the string tension $\mu_l$.
For $g \simeq 10^{-8}$ we enter the range accessible by LIGO, ET, and CE. Further decreasing $g$, we cover the range of DECIGO and LISA. 
Frequencies characteristic to PTAs correspond to extremely small constants $g \lesssim 10^{-18}$. We conclude that gravitational 
radiation from melting cosmic strings serves to probe particle physics in a very weakly coupled regime.

 For substantially larger $f_{\text{peak}}$ and $g$, away from the range accessible by future GW interferometers, emission from melting strings still may have observational consequences through its impact on Big Bang Nucleosynthesis (BBN). GW background acts as dark radiation and thus can be parameterized in terms of the departure from 
 the effective number of neutrino species $\Delta N_{\nu} \approx N_{\nu}-N_{\nu, \text{eff}}$, where $N_{\nu, \text{eff}} \approx 3.046$:
\begin{equation}
 \Omega_{gw, \text{BBN}} \approx \frac{7}{4} \cdot \left(\frac{4}{11} \right)^{4/3} \cdot \frac{\Delta N_{\nu}}{g_* (T_{\text{BBN}})} \; .
 \end{equation}
 Using the Planck bounds $N_{\nu}=2.99 \pm 0.17$~\cite{Planck:2018vyg}, which gives $\Delta N_{\nu} \lesssim 0.11$, and $g_* (T_{\text{BBN}}) \approx 3.4$, we get $\Omega_{gw, \text{BBN}} \lesssim 0.015$. 
 Our prediction of $\Omega_{gw, \text{BBN}}$ can be easily inferred from Eq.~\eqref{peak}, which must be divided by
 \begin{equation} 
 \frac{H^2_{\text{BBN}}}{H^2_0} \left(\frac{a_{\text{BBN}}}{a_0} \right)^4 \approx 2.6 \cdot 10^{-5} \cdot \left(\frac{100}{g_* (T_{\text{BBN}})} \right)^{1/3} \; .
 \end{equation}
 We get 
 \begin{equation}
 \Omega_{gw, \text{BBN}} \simeq 1.7 \cdot 10^{-4} \cdot \left(\frac{G\mu_l}{10^{-4}} \right)^{2} \cdot \left(\frac{g_* (T_{\text{BBN}})}{g_* (T_{l})} \right)^{1/3} \; .
 \end{equation}
The BBN bound on $\Omega_{gw, \text{BBN}}$ can be used to set the upper limit on $G\mu_l$:
\begin{equation}
G\mu_{l} \lesssim 1.6 \cdot 10^{-3}  \; ,
\end{equation}
we have assumed $g_* (T_l) \approx 100$. This is a rather weak bound from the viewpoint of the model~\eqref{modelbasic}, which typically leads to much smaller $G\mu_l$. 
While the future BBN measurements may slightly improve this bound, we expect direct detection of GWs to be the most powerful probe of melting cosmic strings.

\section{Discussions} \label{sec:Section6}
In the present work we have discussed cosmic strings with a decreasing tension~\eqref{decreasing}. Such topological defects can be anticipated if the underlying field theory of particle physics is scale-invariant at high energies (modulo inclusion of gravity), and 
the scale-invariance is not spoiled by quantum corrections. 
In \hyperref[sec:Section2]{Section~2}, we considered an example of a model that enjoys scale-invariance and predicts the existence of melting cosmic strings. We assumed that the latter predominantly 
emit GWs, while the particle emission is negligible. Under this assumption, we estimated the spectrum of GWs and showed that they can be observable 
with future detectors in a well motivated range of parameter space.

Let us summarize our prediction of the spectrum of GWs produced by the network of melting cosmic strings: 
\begin{align}
\label{spectrum}
\Omega_{gw} \cdot h^2_0 \simeq 2.3 \cdot 10^{-9} \cdot \left(\frac{G \mu_l}{10^{-4}} \right)^{2} \cdot \left(\frac{100}{g_* (T_l)} \right)^{1/3} \cdot 
\begin{cases}
\left(\frac{f}{f_{\text{peak}}} \right)^4 \qquad \qquad f \lesssim f_{\text{peak}} \\
\frac{3}{13}  \left(\frac{f_{\text{peak}}}{f} \right)^{1/3} \qquad f \gtrsim f_{\text{peak}}
\end{cases}
\,.
\end{align}
See Fig.~\ref{gw} for further details. In the particular model of cosmic string formation~\eqref{modelbasic}, the peak frequency $f_{\text{peak}}$ is given by Eq.~\eqref{peakfreq}. Let us stress that for $f \gtrsim f_{\text{peak}}$ the expression~\eqref{spectrum} should be viewed as a crude estimate, as our derivation of the GW spectrum relied on some strong assumptions. Namely, we assumed that the cosmic string network immediately 
settles into the scaling regime. Furthermore, we neglected a cosmic string width, which nevertheless may constitute a sizeable fraction of the horizon at string formation time $\tau_l$ (see the discussion in the end of \hyperref[sec:Section2]{Section~2}). We expect that the finite string width strongly affects the GW spectrum at frequencies exceeding $\delta^{-1}_{w}$ at emission. Therefore, it is most possibly not legitimate to extend the expression~\eqref{spectrum} to $f \gg f_{\text{peak}}$. In fact, we expect a strong 
falloff of the spectrum in a very high frequency range.

On the other hand, the expression~\eqref{spectrum} becomes progressively more accurate, as one moves towards low frequencies $f \ll f_{\text{peak}}$. 
The reason is that the low frequency part of the spectrum corresponds to GWs emitted at the times $\tau \gg \tau_l$, when the issues related to the finite string width and settling to the scaling regime are mitigated. In this regard, the spectral shape in the low frequency range, $\Omega_{gw} \propto f^4$, which is also independent of the choice of the loop production function, is the most robust prediction following from our scenario.

In the present work, we did not discuss the cosmological role of the field $\chi$ constituting cosmic strings. According to Eq.~\eqref{peakfreq}, in the range accessible by GWs, the field $\chi$ is very weakly coupled to the thermal bath. Therefore, one could reasonably consider the field $\chi$ to assume the role of Dark Matter. For that purpose, we should admit a small breaking of the scale-invariance by introducing the mass term for the field $\chi$:
\begin{equation}
S_{\text{mass}}=-\int d^4 x\sqrt{-g} \frac{M^2 |\chi|^2}{2} \; .
\end{equation}
Note that for the couplings $g \lesssim 10^{-8}$, which correspond to $f_{\text{peak}} \lesssim 100~\mbox{Hz}$, the standard freeze-out and freeze-in mechanisms are not efficient. Nevertheless, one can create the right amount of Dark Matter at the inverse phase transition~\cite{Ramazanov:2021eya, Babichev:2020xeg, Ramazanov:2020ajq}, 
when the thermal mass of the field $\chi$ drops down to the bare mass $M$. The field $\chi$ gets offset from the minimum around this time and starts 
oscillating. For the masses~\cite{Ramazanov:2021eya} 
\begin{equation}
M \simeq 15~\mbox{eV} \cdot \frac{\beta^{3/5}}{\sqrt{N}}  \cdot \left(\frac{g}{10^{-8}} \right)^{7/5} \; ,
\end{equation}
the energy density of these oscillations matches the observed Dark Matter abundance in the Universe. Despite extremely weak couplings involved, this Dark Matter scenario is naturally connected to 
production of GWs through the early time formation of melting strings, and thus can be tested in future experiments. 

{\it Acknowledgments.} W.~E. and S.~R. acknowledge the Czech Science Foundation, GA\v CR, for financial support under the grant number 20-16531Y. R.~S. is supported  by the  project  MSCA-IF IV FZU - CZ.02.2.69/0.0/0.0/$20\_079$/0017754 and acknowledges European Structural and Investment Fund and the Czech Ministry of Education, Youth and Sports.

\section*{Appendix A: Equation of motion of melting Nambu--Goto strings}\label{sec:AppA}

Dynamics of infinitely thin strings is described by the Nambu--Goto action, which is extrapolated to the case of the time-dependent tension in the straightforward manner~\cite{Yamaguchi:2005gp, Ichikawa:2006rw}: 
\begin{equation}
\label{ng}
S=-\int d^2 \zeta \sqrt{-\gamma} \mu (\tau)\; ,
\end{equation}
where $\zeta =(\zeta^{1}, \zeta^{2})$ are the worldsheet coordinates, $\gamma_{ab} =g_{\mu \nu} x^{\mu}_{,a} x^{\nu}_{,b}$ is the worldsheet metric, and $\gamma \equiv \mbox{det} \gamma_{ab}$. Latin indices stand for the derivatives 
with respect to the worldsheet coordinates. Varying this action and accounting for the time-dependence of the tension, one gets
\begin{equation}
\label{eqmod}
\frac{1}{\mu \sqrt{-\gamma}} \partial_a \left(\mu \sqrt{-\gamma}\gamma^{ab} g_{\mu \nu} x^{\nu}_{,b} \right) -\frac{1}{2} \frac{\partial g_{\lambda \nu}}{\partial x^{\mu}} \gamma^{ab} x^{\lambda}_{,a} x^{\nu}_{,b} -\frac{\partial_{\mu} \mu}{\mu}=0 \; ,
\end{equation}
which can be rewritten in a more conventional form: 
\begin{equation}
\label{convenient}
\frac{1}{\mu \sqrt{-\gamma}} \partial_a \left(\mu \sqrt{-\gamma} \gamma^{ab} x^{\rho}_{,b} \right) +\gamma^{ab} \Gamma^{\rho}_{\lambda \nu} x^{\lambda}_{,a} x^{\nu}_{,b}-\frac{\partial^{\rho} \mu}{\mu}=0 \; .
\end{equation}
Substituting $g_{\mu \nu}=a^2 (\tau) \eta_{\mu \nu}$, where $\eta_{\mu \nu}$ is the Minkowski metric, one can check that the last two terms in Eq.~\eqref{eqmod} cancel each other for $\mu \propto 1/a^2$. 
We used that $\gamma_{ab} \gamma^{ab}=2$. Therefore, for $\mu \propto 1/a^2$, Eq.~\eqref{convenient} simplifies to
\begin{equation}
\partial_a \left(\mu \sqrt{-\gamma}\gamma^{ab} g_{\mu \nu} x^{\nu}_{,b} \right) =0\; .
\end{equation}
We observe that $\mu g_{\mu \nu}=\tilde{\mu} \eta_{\mu \nu}$, where $\tilde{\mu}=\mu a^2=\mbox{const}$, and that $\sqrt{-\gamma} \gamma^{ab}=\sqrt{-\tilde{\gamma}} \tilde{\gamma}^{ab}$, 
where $\tilde{\gamma}_{ab}$ is defined as 
\begin{equation}
\tilde{\gamma}_{ab} \equiv  \eta_{\mu \nu}  x^{\mu}_{,a} x^{\nu}_{,b} \; .
\end{equation}
Finally, we get 
\begin{equation}
\partial_a \left(\tilde{\mu} \sqrt{-\tilde{\gamma}}\tilde{\gamma}^{ab} \eta_{\mu \nu} x^{\nu}_{,b} \right) =0 \; .
\end{equation}
This equation describes evolution of a string with a constant tension in a flat spacetime. In fact, one could anticipate this result from the beginning, based on the scale-invariance of our setup.

Note that Eq.~\eqref{convenient} is generic, as it does not assume a particular time-dependence of the tension $\mu (\tau)$. Here let us make an important comment. A different equation of motion compared to Eq.~\eqref{convenient} has been derived from the same action~\eqref{ng} in Refs.~\cite{Yamaguchi:2005gp, Ichikawa:2006rw}. The latter assume that the tension 
depends on the worldsheet coordinates $\zeta^{a}$ rather than spacetime coordinates $x^{\mu}$ as in our case. Therefore, in Refs.~\cite{Yamaguchi:2005gp, Ichikawa:2006rw}, the variation of the tension $\mu$ is set to zero, when applying the least action principle. However, in our case the tension is directly linked to the temperature of the Universe, which depends on the spacetime coordinates. Consequently, one cannot disregard the variation of $\mu$. 

\section*{Appendix B: VOS model} \label{sec:AppB}

In the VOS model, the loop production function is given by Eq.~\eqref{onescale}. In the present Appendix, we aim to analytically derive 
the coefficient $C$ entering there. Our discussion of the VOS model mainly follows Refs.~\cite{Martins:1996jp, Martins:2000cs}. 

One assumes that loops are produced from long strings according to
\begin{equation}
\label{loss}
\frac{d\tilde{\rho}_{\infty}}{d\tau} =-\bar{c} v \frac{\tilde{\rho}_{\infty}}{\tilde{L}} \; ,
\end{equation}
where 
\begin{equation}
\label{longenergy}
\tilde{\rho}_{\infty}=\frac{\tilde{\mu}}{\tilde{L}^2} \; ,
\end{equation}
is the rescaled energy density of long strings; $\bar{c}$ is the so-called loop chopping efficiency parameter defined numerically, and $v$ is the root mean square velocity. 
Comparing Eqs.~\eqref{loss} and~\eqref{longenergy}, we obtain the scaling behaviour of long strings: 
\begin{equation}
\label{longscaling}
\frac{\tilde{L}}{\tau} =\frac{\bar{c} v}{2} \; .
\end{equation}
Flat spacetime simulations of Refs.~\cite{Martins:2003vd, Moore:2001px} give 
\begin{equation}
\label{chopping}
\bar{c} \approx 0.57 \; .
\end{equation}
The velocity $v$ is obtained from the equation~\cite{Martins:1996jp, Martins:2000cs}  
\begin{equation}
\frac{d v}{d \tau}=(1-v^2) \cdot \frac{k (v)}{\tilde{L}} \; , 
\end{equation}
where the function $k (v)$ is given by
\begin{equation}
k (v)=\frac{2\sqrt{2}}{\pi} (1-v^2) \cdot (1+2\sqrt{2} v^3 ) \cdot \frac{1-8 v^6}{1+8 v^6} \; .
\end{equation}
The system has an attractor solution 
\begin{equation}
\label{velocity}
v =\frac{1}{\sqrt{2}} \; ,
\end{equation} 
as it should be in a flat spacetime~\cite{Vilenkin}. 

The energy loss~\eqref{loss} is related to the loop production function $f(x)$ by
\begin{equation}
\label{energyloss}
\frac{d\tilde{\rho}_{\infty}}{d\tau} =-\tilde{\mu} \int^{+\infty}_0 \tilde{l} f(\tilde{l}, \tau) d\tilde{l} \; .
\end{equation}
Using Eqs.~\eqref{loss} and~\eqref{longscaling}, we get
\begin{equation}
\label{intvos}
\int^{+\infty}_0 x f(x) dx =\frac{8}{\bar{c}^2 v^2} \; .
\end{equation}
Substituting Eq.~\eqref{onescale}, we obtain
\begin{equation}
C=\frac{8}{\alpha \bar{c}^2 v^2} \; .
\end{equation}
Finally, substituting the values~\eqref{chopping},~\eqref{velocity} and using $\alpha \simeq 0.1$, we obtain 
\begin{equation}
C \simeq 500 \; .
\end{equation}
This is in a good agreement with the value~\eqref{numerics} derived from matching to numerical simulations of Ref.~\cite{Vanchurin:2005pa}.

\end{document}